# LA ORIENTACIÓN DE LAS IGLESIAS ANDINAS DE LA REGIÓN DE ARICA Y

# PARINACOTA, CHILE: UNA APROXIMACIÓN ARQUEOASTRONÓMICA

*Alejandro Gangui* [1], *Ángel Guillén* [2] y *Magdalena Pereira* [3]

1. Doctor en Astrofísica por la International School for Advanced Studies, Italia, investigador de CONICET y profesor de la Universidad de Buenos Aires, Argentina. Correo electrónico: algangui@gmail.com

2. Magister en Desarrollo Sustentable por la Universidad de Lanús, Argentina, jefe del taller de arquitectura y patrimonio de la Fundación Altiplano, Chile. Correo electrónico: angel.guillencardenas@gmail.com

3. Doctora en Historia del Arte por la Universidad de Sevilla, España, directora e investigadora de la Fundación Altiplano, Chile. Correo electrónico: pereiramagdalena@gmail.com

**Resumen**

Las iglesias patrimoniales de Arica y Parinacota son representativas de la mayoría de los templos cristianos de la región. A juzgar por su arquitectura, materialidades y su decoración, tienen características en común y son fieles a las intenciones de sus constructores originales. Sin embargo, sus emplazamientos geográficos y el paisaje que las rodea guarda características propias para cada una de ellas. En particular, la orientación de sus ejes principales (en dirección al altar de las iglesias) muestra una gran diversidad. En esta extensa región, con escasa atención de doctrineros, la identidad y realidad aymara local, con su cosmovisión, debieron dialogar con la tradición occidental para configurar y construir los pueblos de las reducciones indígenas. En este trabajo se presentan los resultados obtenidos a partir del análisis de las orientaciones espaciales precisas de la totalidad de las iglesias catalogadas, empleando las herramientas de la arqueoastronomía, como complemento a la investigación y análisis arquitectónico, histórico y cultural de los templos. Mencionamos también algunos avances que esperamos poder realizar en zonas vecinas a la región, las que incluyen un número grande de templos antiguos y forman parte, junto a las iglesias aquí estudiadas, de una común y extensa *koine* cultural.

**Palabras clave:** iglesias andinas de Arica y Parinacota, arqueoastronomía, orientación de iglesias.

**Abstract**

The heritage Andean churches of Arica and Parinacota are representative of the majority of Christian churches in the region. Judging by their architecture, materiality and decoration, they have many features in common and are faithful to the intentions of their original builders. However, their geographical locations and the landscape that surrounds these churches show specific features for each of them. In particular, the orientation of the main axes (towards the altar of the churches) shows some diversity. In this extended area, with little attention from parish priests, the local Aymara culture, with its own worldview, surely engaged in a dialogue with the Western tradition in order to design and build the Indian Reductions. In this paper we present the results obtained from the analysis of the precise spatial orientations of all the cataloged churches, employing the tools of archaeoastronomy, as a complement to the architectural, historical and cultural research of the temples. We also comment briefly on some progress that we hope to make in neighboring areas to the region here studied, which include a large number of very old temples and are part, along with the churches here considered, of a common and widespread cultural *koine*.

**Introducción**

La orientación espacial de las iglesias cristianas antiguas es uno de los elementos distintivos de su arquitectura. En Europa y en muchos sitios lejanos a donde llegó la evangelización, existe una tendencia general a orientar los altares de los templos en el rango solar, es decir, orientarlos hacia aquellos puntos del horizonte por donde sale el Sol en diferentes días del año, con una clara predilección por las orientaciones cercanas al este geográfico (equinoccio astronómico) (McCluskey 1998:26; González-García y Belmonte 2015). Dentro del mismo rango solar, sin embargo, no son inusuales las alineaciones en sentido opuesto, con el altar a poniente, aunque resultan excepcionales pues no siguen el patrón canónico (Esteban et al. 2001; Belmonte et al. 2007).

Poco tiempo después del asentamiento español en el virreinato peruano, el virrey Francisco de Toledo reorganiza el territorio y su población (Toledo 1989:450), poniendo especial enfoque en las rutas comerciales en el ámbito surandino, cuyo principal objetivo fue el de procurar la salida de la plata desde Potosí al Pacífico y del azogue desde Huancavelica a las minas alto-andinas. Camino hacia Arica, puerto oficial de salida de la plata y entrada del azogue desde 1574, se formaron pequeños caseríos y tambos con poblaciones estables. Las iglesias andinas de esta región surgieron en sitios estratégicos a lo largo de la ruta que trajinantes recorrían para transportar los metales preciosos desde Potosí hacia las playas de Arica, en particular en torno a los valles de Lluta y Azapa. Al ser una zona extensa y difícil de transitar, la influencia de los doctrineros en la configuración de los poblados y reducciones indígenas fue limitada. Este hecho acentuó el diálogo entre las culturas locales y la tradición occidental, y esta interacción pudo ser el origen de muchas características propias de la arquitectura andina; las iglesias construidas en épocas coloniales forman un corpus de estudio importante para intentar develar algunos elementos de dicha interacción.

En este trabajo estudiamos un conjunto grande de iglesias patrimoniales de la región, concentrando nuestra atención en sus orientaciones, emplazamientos geográficos y en el paisaje que las rodea. Medimos la orientación de un total de 38 iglesias andinas, ubicadas en el altiplano, sierra y valles bajos de las quebradas de Lluta, Azapa, Vítor y Camarones. Las dos primeras por donde bajaba la ruta de la plata; las dos últimas, fértiles oasis funcionales a proveer a los pueblos y tambos del sustento necesario para los trajines. La mayoría de estas iglesias fueron reconstruidas durante el siglo XIX, con algunos pocos ejemplos de épocas anteriores y posteriores. Como veremos, la muestra indica que, aunque no se siguió un único patrón de orientación determinante en toda la región, casi la mitad de las iglesias estudiadas se orienta dentro del rango solar, con una proporción dominante en aquellas que presentan su altar hacia el poniente.

En las próximas tres secciones nos concentramos en el contexto histórico y cultural de la región estudiada y en el rol importante que en ésta ha desempeñado la ruta de la plata. En la última parte de nuestro trabajo detallamos los estudios arqueoastronómicos llevados a cabo en lo referente a las orientaciones de los templos de los Altos de Arica y analizamos nuestros resultados. Finalmente, en las conclusiones discutimos algunos de los elementos que podrían arrojar luz sobre los patrones de orientación encontrados en nuestras mediciones y sobre sus causas, y comentamos brevemente futuros proyectos de extender nuestro trabajo a conjuntos de iglesias antiguas de regiones vecinas.

**Contexto histórico y cultural de la región**

El territorio de Arica y Parinacota ha sido desde tiempos pretéritos una región de alto interés por su emplazamiento en un punto estratégico del continente, donde altiplano y costa se encuentran íntimamente ligados, con una gran diversidad natural y cultural. Aquí se distinguen tres grandes pisos ecológicos: costa y valles bajos, sierra o precordillera y puna o altiplano en sólo 16.800 km$^2$.

Los valles bajos, situados entre los 0 y los 1.900 m.s.n.m., desafían al árido desierto con cursos de agua originados en las cumbres del altiplano y las vertientes de la sierra, los que generan cuencas y reservas naturales. La región presenta, de norte a sur, cuatro cuencas principales: las ya mencionadas Lluta, Azapa, Vítor y Camarones, que posibilitan la existencia de variada flora y fauna. Su morfología de cerros y quebradas que se originan en la cordillera de los Andes, con altitudes entre los 2.000 y los 3.500 m.s.n.m., ha sido escenario de importantes desarrollos culturales (Cunill Grau 2014; González Jiménez 2014). En la parte media de la extensa cordillera de los Andes se genera una gran planicie, el altiplano. Con una altura promedio de cuatro mil metros, comprende el sur de Perú, Bolivia occidental, el norte de Chile y el noroeste de la Argentina. Los lagos Titicaca y Poopó, en la meseta del Collao, riegan la zona alto-andina, permitiendo la sobrevivencia en un clima hostil. El Titicaca, de agua dulce, con 8.562 km$^2$ de superficie y ubicado a una altura de 3.812 m.s.n.m., desagua en el lago Poopó a 3.690 m.s.n.m. Al sur se encuentran los lagos salados y el salar de Coipasa, donde rebalsa el Poopó y el Uyuni, este último independiente (Bouysse-Cassagne 1987:35).

En esta rigurosa geografía, durante el período Intermedio Temprano o también período Formativo, aproximadamente entre los años 1000 a.C. y 500 d.C. (Muñoz 2004), habitaron grupos alto-andinos que domesticaron los camélidos y que se desplazaban en caravanas de llamos conectando el altiplano y la costa. Durante el período Medio, aproximadamente entre los años 500 y 1000 d.C., el imperio *Tiwanaku* expandió su cultura desde el lago Titicaca hacia la costa. Los Desarrollos Regionales, aproximadamente en el período 1000-1400, surgen con la desintegración de *Tiwanaku*; entre ellos, la cultura Arica, época de la cual datan los geoglifos de los valles de Lluta y de Azapa. Hacia 1470, con el período Inca se unifica el vasto mundo surandino.[1] De la influencia de *Tiwanaku* quedan etnias altiplánicas habitando el altiplano peruano, boliviano y chileno, practicando el intercambio con la sierra y valles costeros: Pacajes, Lupacas, Carangas y Uros, entre otras. Estos grupos familiares, organizados en ayllus en el período incaico, sobrevivirán en el desierto complementando su dieta en los distintos pisos ecológicos de la costa, valles bajos, precordillera y altiplano y se adaptarán a la conquista inca y española resignificando su cosmovisión (Murra 2009:85-86).

El descubrimiento del mineral de Potosí en 1545, marcó la inflexión en la economía del virreinato peruano. Como señalamos, el virrey Toledo se encargó de fomentar la explotación del mineral, reorganizando el sistema de trabajo incaico, la mita. Durante dos siglos y medio los antiguos corredores andinos fueron utilizados como camino real o de la plata, y cientos de miles de llamos y mulares cargaron la plata y el azogue desde Huancavelica, con todo lo necesario para surtir a la villa imperial de Potosí, dirigiéndose hacia Arica, puerto oficial de entrada y salida de la mercadería (Moreno y Pereira 2011:37). El 17 de Julio de 1565 el virrey Toledo decretó la creación del Corregimiento de Arica, con las reparticiones de Tarapacá y Pica, y el Corregimiento de Lluta con sus reparticiones de Arica, Ilo, Ite, Ilabaya y Tacna (Sanhueza Tohá 2008). Arica contó con cajas y almacenes reales donde se pagaban los impuestos y donde eran almacenados los minerales del alto Perú (Málaga Medina 1989:167-168).

**La iglesia en la región y la orientación de los templos**

En términos eclesiales, en el año 1537 se creó el obispado de Cuzco, que comprendió una extensa zona del altiplano peruano, boliviano y chileno. Luego, en 1613 con la creación del Obispado de Arequipa, Arica y sus anexos pasaron a depender de éste. En 1618, el sacerdote carmelita Antonio Vázquez de Espinosa menciona el repartimiento de Lluta y Arica como doctrinas atendidas por clérigos que tenían en esa época 65 tributarios, 15 viejos, 84 muchachos y 92 mujeres (Vázquez de Espinosa 1948 [1629]:655-657; Marsilli y Cisternas 2010). Lluta fue la primera doctrina en el siglo XVI de la cual dependió la atención de los pueblos del interior, la que luego fue traspasada a Azapa en el siglo XVII y a Codpa en 1668. Este pueblo pasó a ser cabecera de doctrina, con anexos en los pueblos del interior: en Tímar, Esquiña, Pachica, Ticnámar, Saxámar, Belén, Socoroma, Putre, Guallatire, Choquelimpie, Pachama, Parinacota, Caquena, Livilcar y Humagata.

En Arica se asentaron las órdenes religiosas de San Juan de Dios, hospitalarios, jesuitas, franciscanos y mercedarios. Mantuvieron convento en la ciudad y destinaban a religiosos como colaborador o teniente de cura a las doctrinas del interior. Los franciscanos tuvieron además en la chimba de Arica una hospedería, que funcionaba en 1636 para los misioneros que iban de paso a evangelizar a la sierra, a la ruta de la plata. Durante los siglos XVII y XVIII era común en las doctrinas del nuevo mundo que los religiosos se hicieran cargo de las doctrinas de indios o ayudaran a los curas diocesanos en su tarea, en las zonas rurales y alejadas (Hidalgo y Díaz 1985:82).

A la construcción de las dependencias reales del nuevo puerto de la plata, se sumaron las reducciones para indígenas que el virrey Toledo decretó hacia 1570. En base a esto, los nativos debían reunirse en pueblos para ser evangelizados de manera más eficiente, ocasión en que se comenzó la construcción de los primeros templos andinos. Es de suponer que las primeras iglesias construidas en la región estuvieron bajo la supervisión de los curas evangelizadores. Interesa indagar qué alcance tuvieron las tradiciones arquitectónicas traídas por estos misioneros al construirse los templos cristianos en tierras y paisajes muy diferentes de aquellos del viejo mundo.

Según los textos de los primeros escritores cristianos, las iglesias debían construirse siguiendo una determinada orientación: aquella que asegurase que el sacerdote se situaba mirando hacia el oriente durante los oficios. Así lo reconocen, entre otros, Orígenes, Clemente de Alejandría y Tertuliano, y fue avalado en el año 325 por el Concilio de Nicea. San Atanasio de Alejandría, también en el siglo

IV, expresa que el sacerdote y los participantes deben dirigir su mirada hacia el este, de donde Cristo, el Sol de Justicia, brillará al final de los tiempos (Vogel 1962).

En el siglo V, Sidonio Apolinar y Paulino de Nola indican que el ábside de las iglesias debe mirar hacia el este, al equinoccio, algo confirmado más tarde tanto por el Papa Virgilio como por Isidoro de Sevilla en sus *Etymologiae* (XV, 4; McCluskey 1998:125). Esto sería confirmado durante la Edad Media por Honorio Augustodunense (siglos XI-XII) y por otros autores como Guillermo Durando (siglos XII-XIII), que indican claramente la dirección a seguir, el equinoccio, y aquella a evitar, el solsticio (McCluskey 2004, 2010). Esto último, la necesidad de alejar las orientaciones de los templos de las direcciones cercanas a los solsticios, podría estar ligado a la importancia de estas fechas en períodos anteriores al cristianismo y a los numerosos templos paganos construidos con esas orientaciones.

Tengamos en cuenta, sin embargo, que en las prescripciones ya señaladas todavía persiste un tanto de ambigüedad al orientar las iglesias hacia el este, pues cabría preguntarse hacia qué equinoccio hacerlo. Como menciona McCluskey (2004) existen varias posibilidades: el equinoccio vernal romano ocurría el 25 de marzo, mientras que el griego estaba fijado para el 21 de marzo, como quedó plasmado en el Concilio de Nicea. Sin embargo, también podían usarse otras definiciones, tales como la entrada del Sol en el signo de Aries o el equinoccio de otoño. En los amaneceres de cada una de estas fechas diferentes el Sol se hallaría en sitios distintos del horizonte y, por lo tanto, las orientaciones resultantes de los templos serían diferentes (Ruggles 1999; González-García y Belmonte 2006). Por otra parte, también debemos considerar el uso del Calendario Juliano durante la Edad Media y buena parte de la Moderna. La naturaleza de éste haría que, si nos fijamos en un equinoccio calendárico (es decir en una fecha concreta), tal momento se desplazaría con respecto a las estaciones, algo que se vería reflejado en un cambio sistemático de orientación de las iglesias, si ésta se hacía por observación de la salida del Sol en ese día.

La influencia más directa para la construcción de templos en nuestra área de estudio la encontramos en el siglo XVI, luego del Concilio de Trento (1545-1563), en que el Cardenal Carlos Borromeo publica sus *Instrucciones de la fábrica y del ajuar eclesiásticos* (traducidas y publicadas por Bulmaro Reyes Coria en México en 1985), con gran difusión en la época. En estas señala la dirección que debe tener el altar mayor:

> Ahora bien, el sitio de esta capilla debe elegirse en la cabeza de la iglesia, en el lugar más elevado por cuya región esté la puerta principal; su parte posterior mire en línea recta hacia el oriente, aunque los domicilios del pueblo estén por la parte de atrás. Y no se sitúe nunca completamente hacia el oriente solsticial, sino hacia el equinoccial (…) Esta capilla esté abovedada; además adórnese con decoro mediante obra de mosaico, o mediante otra especie de pintura o estructura ilustre, según el tipo y dignidad de la iglesia que se edifique (Borromeo 1985:15).

Sabemos que las reducciones tardaron en llevarse a cabo; Toledo se queja de esto en 1575, y nombra en sus disposiciones a examinadores o responsables que las ejecutasen, incluyendo a clérigos en esta misión. Toledo dejaba en claro que se debían destruir los pueblos antiguos (de hecho, Vázquez de Espinosa (1948) escribe: "quemé un pueblo que se llamaba Isquiliza, porque los más eran idólatras"), y que las nuevas poblaciones debían construirse a una distancia prudente de las antiguas edificaciones (Toledo 1989:415).

Las prospecciones arqueológicas en los pueblos de los templos en estudio han establecido que, al menos en los pueblos principales, en cierto sentido se cumple esta condición. En tanto, las ruinas que señalan poblamientos prehispánicos se encuentran en los alrededores y a una distancia "prudente" de los pueblos principales, tales como Codpa, Belén y Socoroma. Excepción a esto es quizás el poblado de Saguara en que el templo fue construido a fines del siglo XIX con las piedras de un *ushnu* incaico ubicado a solo cincuenta metros de allí, y orientado con precisión en forma casi perpendicular (acimut 266.5°) a la antigua construcción (acimut 354.5°), como pudimos verificar en nuestra visita.[2] En el caso de los templos restaurados, Socoroma por ejemplo, paso obligado en el Qhapaq Ñan y en el camino de la ruta de la plata, fueron encontrados restos cerámicos de los distintos períodos de poblamiento (Tiwanaku, cultura Arica, Inca y colonial). Sin embargo, en Socoroma no se hallaron indicios de que este fuese previamente un recinto habitacional, es decir, no es posible establecer que la iglesia haya sido construida sobre antiguas huacas o sobre un pueblo prehispánico (Pereira 2013:131-135).

**Cambios políticos y sociales. El arraigo a las costumbres religiosas**

La política de los Borbones en el siglo XVIII debutó con una excesiva carga de impuestos y, a la par, comenzó a oscilar la producción de plata de Potosí. Estos eventos fueron lentamente provocando una tendencia a la baja del esplendor del puerto de Arica y de los pueblos del interior. Sumado a esto, la instalación de las cajas reales en Tacna, en 1717, provocó la migración de importantes vecinos al valle aledaño y, posteriormente, en la década de 1780, Arica pasó a constituir un partido dentro de la Intendencia de Arequipa, lo que contribuyó a la decadencia de la ciudad. Recordemos que la creación del virreinato del Río de la Plata con capital en Buenos Aires, en 1776, ya había desviado la ruta del mineral de Potosí hacia el Atlántico, acentuándose la situación de decaimiento del antiguo puerto de la plata (Moreno y Pereira 2011:40-41).

Durante el siglo XIX, el territorio de Arica y Parinacota fue escenario de acontecimientos relevantes: las luchas por la independencia del Perú, que se logra en 1821; la Guerra de la Confederación peruano-boliviana, 1836-1839; los terremotos de 1868 y 1877; y finalmente la Guerra del Pacífico en 1879. Todos estos eventos obligaron a los habitantes y autoridades a levantar una y otra vez la derruida ciudad. En particular, el fuerte terremoto de 1868 provocó una importante destrucción en el centro histórico y varios templos del interior. Entre los edificios colapsados se encontraban la Aduana antigua (inaugurada hacia 1855) y la Basílica Parroquial de San Marcos (terminada en 1640). Entre las obras nuevas se encontraba la Iglesia Matriz, actual Catedral de San Marcos (Revista del Sur 1870:2). Al nuevo templo se le menciona como la "iglesia de fierro" (supuestamente prefabricado en Francia) e instalado en Arica en el año 1874.

Por su parte, las comunidades andinas junto al clero peruano no se mantuvieron ajenos a la reconstrucción; activos en el territorio ocupado por Chile, sacerdotes del Perú atendían a la feligresía en Arica, Tacna y en las parroquias rurales del interior. En los archivos parroquiales de la época se registran trabajos en muchas iglesias andinas, que deben ser incluso reconstruidas completamente, debido a los graves daños provocados por los sismos de 1868, 1877 y 1906, tales como Poconchile, Azapa, Tignámar, Putre y Socoroma, por señalar algunas. En todo este tiempo, el Arzobispado de Arequipa mantuvo sus títulos de propiedad y sus sacerdotes hasta 1910, cuando el Estado chileno resolvió expulsar al clero peruano y expropiar las propiedades, para evitar todo tipo de soberanía peruana en la zona ocupada. Fue así el gobierno chileno construyó en el área de Arica escuelas destinadas a generar conciencia nacional chilena en los estudiantes, donde los profesores actuaban como verdaderos promotores de la identidad y soberanía patriótica. Por su parte, es lógico pensar que el clero peruano se mantuvo cercano a la feligresía a través de la reconstrucción y alhajamiento de sus templos históricos, que constituían y constituyen aun en nuestros días el eje central de la vida en comunidad y que representan el corazón de sus tradiciones y costumbres (Moreno y Pereira 2011:87-88).

**Arqueoastronomía de las iglesias andinas**

En el siglo XIX la actividad misional en los poblados de la región de Arica y Parinacota se

concentraba en los aleros de las parroquias. La organización y construcción de capillas estaba en

manos de los curas párrocos y de las comunidades andinas. El marcado carácter conservador de

estos protagonistas de la vida religiosa hizo que las instituciones y ritos coloniales se mantuvieran

con celo en los templos, haciendo de estos los ejes principales de su vida comunitaria. Las

celebraciones patronales cobraron una importancia sustantiva, acogiendo los ritos de la religiosidad

andina ancestral en las "costumbres" que se celebran hasta el día de hoy en los atrios de las iglesias

(Díaz et al. 2012).

El entrelazamiento simbólico cultural permanece grabado en los templos con huellas notables.

Testimonios del pasado, los elementos y símbolos ornamentales europeos conviven con los autóctonos: las vizcachas talladas en las columnas salomónicas de la portada de algunas iglesias,

como la de Livílcar, o el mono, el hombre-puma y las aves nativas presentes en las columnas de la

portada de la iglesia de Santiago Apóstol en el poblado de Belén, junto a pinturas murales que

muestran la imagen de un trifronte en la escena de las postrimerías de la iglesia de Parinacota, son

algunos ejemplos representativos.[3]

El estudio y análisis del conjunto patrimonial de iglesias andinas de esta región es un capítulo fundamental en las historias del arte y de la iglesia; también lo es en la arquitectura y en la arqueología del paisaje andino. En esta parte nos concentraremos en el estudio de los emplazamientos de los templos en el paisaje de la región, poniendo especial énfasis en analizar la posibilidad de que sus orientaciones hayan sido dictadas por elementos del entorno terrestre, como cerros prominentes o volcanes, o celeste, por ejemplo si los ejes principales de las iglesias fueron elegidos en coincidencia con los sitios del horizonte por donde sale el Sol en algún día especial del año.[4]

Para este estudio es fundamental contar con un grupo numeroso de construcciones, de tal manera de poder llevar a cabo un trabajo estadístico de búsqueda de correlaciones. En base al elevado número de monumentos históricos patrimoniales documentados, cercano a los cuarenta y por lo tanto estadísticamente significativo, pensamos que nuestras mediciones pueden ser representativas de la mayoría de los templos cristianos antiguos de la región. Nuestro proyecto a largo plazo es indagar, además, si la construcción de las iglesias de la región fue influenciada por factores tales como la presencia de la cultura y población aymara local que, en principio, fue heredera de unos patrones de culto muy diferentes a los de los colonizadores (Gisbert 1999:4-6; Díaz et al. 2012).

La Tabla 1 muestra los datos obtenidos en una campaña de trabajo de campo llevada a cabo en marzo de 2015. Se presentan las cantidades estándar (identificación de la iglesia o capilla y sus coordenadas) junto con las referencias de su orientación (datos arqueoastronómicos): acimut y altura angular del horizonte (medidos), y la declinación astronómica correspondiente (calculada).[5] En lo referente a las fechas consignadas para la construcción de las iglesias, existe cierta ambigüedad, pues las fuentes documentales son escasas y, muchas veces, se refieren a iglesias antiguas en la zona que ya no están en pie.

Obtuvimos las mediciones empleando un tándem Suunto 360PC/360R/D (precisión 0.5°), calibrado para el hemisferio sur (zona 4), que incorpora un clinómetro y una brújula de alta precisión, analizando asimismo el entorno paisajístico de cada uno de los edificios.[6] Posteriormente corregimos los datos de acuerdo a la declinación magnética local. Nuestros valores de la declinación magnética para distintos sitios de la región oscilan dentro del rango de 1° (desde 5°05' a 6°05' oeste). Como la precisión que tenemos con los acimutes magnéticos medidos es de aproximadamente 1/2°, en lugar de considerar un valor promedio para toda la muestra hemos corregido cada medición con la declinación magnética correspondiente de su sitio.

Los valores establecidos para los acimutes resultan de la combinación de varias mediciones que toman en cuenta las orientaciones no solo de los ejes de las iglesias, desde la puerta hacia el altar, medidas siempre que pudimos acceder a sus interiores, sino también las orientaciones de los muros, testero y de la fachada, y de los laterales de los edificios, siempre que estos no fueran demasiado irregulares. Debemos enfatizar que, en general, las diferentes mediciones resultaron en una orientación para cada iglesia con una incerteza de menos de 1/2°.

Para confirmar nuestros datos obtenidos in situ, y dadas las alteraciones magnéticas consignadas en varios lugares de la región (sobre todo en los alrededores de los imponentes volcanes presentes en ella), se han verificado algunas medidas con imágenes fotosatelitales, encontrándose pocas divergencias.[7] Por ello, estimamos que el error de nuestras medidas está en torno al 1/2° como cota superior, siendo así los datos lo suficientemente precisos como para abordar el estudio estadístico deseado.

En la Figura 1 se muestra el diagrama de orientación para las iglesias y capillas patrimoniales estudiadas. Como hemos mencionado, los valores de los acimutes son los medidos en cada sitio, e incluyen las correcciones por declinación magnética. Las líneas diagonales del gráfico señalan los acimutes correspondientes −en el cuadrante oriental− a los valores extremos para el Sol (acimutes de 65.4° y 115.0° −líneas continuas−, equivalente a los solsticios de invierno y verano austral, respectivamente) y para la Luna (acimutes: 59.2° y 120.7° −líneas rayadas−, equivalente a la posición de los lunasticios mayores o paradas mayores de la Luna).

De las 38 iglesias medidas, seis se encuentran orientadas hacia el cuadrante norte (es decir, entre acimutes -45°, o bien 315°, y 45°). Hay ocho orientadas en el cuadrante a levante (seis de ellas en el rango solar)[8] y 16 orientadas en el cuadrante a poniente (11 de ellas en el rango solar). Finalmente, hay ocho iglesias en el cuadrante meridional (entre acimutes 135° y 225°). Entendemos que nuestra muestra es representativa de la mayoría de las iglesias antiguas de la región. En nuestros datos se distingue una orientación destacada, hacia el cuadrante occidental, con 16 iglesias que se alinean dentro de ese rango de acimutes. Entre estas, hay 11 iglesias que entran dentro del rango solar, entre los acimutes 245.0° y 294.6°.

Llegados a este punto, debemos tener en cuenta que en muchos de los sitios estudiados las condiciones geográficas son singulares y seguramente han jugado un papel no menor en el emplazamiento de las iglesias. Tal es el caso de la iglesia de Chitita, situada en el cañón del río Vítor y rodeada de montañas bajas. En el sitio de su emplazamiento, la pendiente del terreno perpendicular a su eje va a dar directamente hacia el lecho del río. El sendero que pasaba por ese sitio y que luego sirvió para acceder a la iglesia, muy probablemente iba copiando ese cauce y la disposición más natural de la construcción no podía apartarse demasiado de la actual (que posee un eje alineado con 299.5° de acimut). De ser esta la verdadera razón de su orientación, y luego de observar el paisaje circundante, se infiere que, de haber sido construida unos 500 metros más hacia el oeste, y siempre a lo largo del río, su orientación podría haber cambiado en más de 40 grados de acimut. Algo similar ocurre con la iglesia de Aico, ubicada en el valle interior de la quebrada de Atco. Aquí también la orientación (69.5° de acimut) parece seguir aproximadamente la traza de un río.

Un caso diferente es, por ejemplo, la iglesia San Antonio de Padua, en el poblado de Sucuna, ubicada en una planicie muy cercana a la quebrada de Sacuna pero aparentemente no condicionada por elementos topográficos. En este caso, la iglesia se orienta con un acimut de 275.5°, su eje interseca el horizonte a pocos grados del lugar de la puesta del Sol en los equinoccios y su declinación calculada entra bien en el rango solar. Algo similar sucede con la ya mencionada capilla de Saguara, orientada también hacia el poniente (266.5° de acimut), y con la iglesia del poblado de Pachica, ubicada sobre una colina que desciende hacia la quebrada de Saguara, la que a su vez lleva a la quebrada de Camarones, pero en este caso con el eje de la construcción orientado hacia el levante (76° de acimut).

Para intentar comprender mejor lo expuesto, en la Figura 2 presentamos el histograma de declinaciones, que resulta independiente de la ubicación geográfica y de la topografía local. Éste muestra la declinación astronómica frente a la frecuencia normalizada, lo que permite una más clara y acertada determinación de la estructura de picos.

Como en trabajos anteriores (por ejemplo, Gangui et al. 2016), aquí hemos empleado una función "kernel" de suavizado apropiada para estimar la declinación, lo que da origen a una "distribución de densidad kernel" (que notamos DDK) caracterizada por un paso de banda (en nuestro análisis empleamos un kernel Epanechnikov con 2° de paso de banda). Así, para cada valor de declinación, se multiplica ese valor por la DDK con el paso de banda elegido. Luego, todos estos productos (DDKs) se suman para dar la DDK final de nuestros datos. Para poder afirmar que una medición es significativa, empleamos una frecuencia relativa normalizada que fija la escala de nuestras DDKs o histogramas: se divide el número de ocurrencias de un dado valor por el valor promedio de ocurrencias de la muestra. Esto es equivalente a comparar con el resultado de una distribución uniforme en declinación del mismo tamaño que nuestra muestra de datos y con un valor igual al del promedio de nuestros datos. Se espera que al graficar los datos así procesados surjan picos en ciertos valores de declinación con amplitud por encima de cero, idealmente por encima del valor 3 (llamado 3σ).

En el caso de iglesias medievales europeas (González-García, 2015),[9] gráficos de este tipo muestran un pico muy pronunciado sobre el valor cero de declinación, aunque con un cierto ancho, lo que indica que existe una cierta dispersión en las orientaciones, pero que la mayoría de esas iglesias sigue la dirección del equinoccio astronómico. Eso no sucede con nuestros datos, donde vemos pequeñas acumulaciones de orientaciones en un rango muy extenso de declinaciones, muy pocas estadísticamente significativas.

Por otra parte, en lo referente a construcciones religiosas, pequeñas y visibles, que rodean a las iglesias –como calvarios y apachetas con cruces, aunque no todos antiguos–, hemos encontrado muchos ejemplos representativos. Tal es el caso de la iglesia de Aico, rodeada por al menos dos calvarios, ubicados en sendas colinas y aproximadamente alineados con su eje. Una de estas estructuras se ubica detrás de la iglesia en dirección 66° de acimut, mientras que la otra, en dirección 247° de acimut, es visible al salir del interior del templo. Otro caso notable es el de la iglesia de Parcohaylla, rodeada por cuatro de estas pequeñas construcciones, muchas veces con senderos bien marcados para que los fieles puedan acceder a ellas.

Aparte de los cauces de ríos y quebradas ya mencionados, los numerosos volcanes y picos nevados de la región pueden servir de referencia –como cerros tutelares o apus (Reinhard 1983), incluso relacionados con el culto a los ancestros (Bouysse-Cassagne y Chacama 2012)–[10] a la hora de decidir el emplazamiento y la orientación de los templos, por lo que conviene inspeccionar nuestros datos para verificar esa posible orientación orográfica.

Son bien conocidas las vinculaciones que existen entre las comunidades andinas y los espacios sagrados de los Andes, como los mencionados apus o las achachilas (que se refiere a "abuelo" o "antepasado"), de suma importancia para la vida cultural y religiosa de los habitantes de la zona (Leoni 2005; Martínez 1983; Van den Berg 1989; Van Kessel 2001). Trabajos etnográficos recientes, pese al conocido despoblamiento de la región, han revelado interesantes conexiones entre las comunidades y el culto a los cerros (Choque y Pizarro 2013), elementos que han perdurado en el tiempo ligados a su cosmovisión.

En el caso de la capilla Virgen del Carmen de Ungallire, al salir de su interior, en el frente, uno se encuentra con la visión imponente de los payachatas (los volcanes Pomerape y Parinacota); estos se ubican a solo unos pocos grados a uno y otro lado del eje principal de la construcción, como puede verse en la Figura 3.[11] Se verifica algo similar con las iglesias Virgen de la Inmaculada Concepción de Guallatire y de Ancuta: de acuerdo a nuestras mediciones, el volcán Guallatiri se encuentra casi alineado con los ejes de estas iglesias pero, en ambos casos, y como sucedía con la capilla de Ungallire, la ubicación del volcán está en el frente de las construcciones.

Sin embargo, esto es muy diferente de lo que sucede con otras construcciones, por ejemplo, con la iglesia de Tacora; aunque ésta tiene al volcán del mismo nombre muy cerca de su emplazamiento, la línea de su eje difiere en unos 30° respecto de la dirección en que se halla dicha elevación prominente del paisaje. Lo mismo encontramos con la iglesia de Parinacota y su volcán homónimo: el eje de la iglesia y la dirección hacia el volcán difieren en más de 30° de acimut, por lo que la presencia de los volcanes no parece haber sido un factor influyente al momento de orientar estos templos.

En base a este análisis, pensamos que, aunque no siempre, en muchas ocasiones (por ejemplo, en los mencionados templos de Chitita, Aico, Ungallire, Guallatire y Ancuta) las características propias –la topografía, el paisaje circundante– de cada sitio donde fueron instaladas las iglesias andinas primaron sobre las tradiciones europeas y coloniales en lo referente a la orientación de sus ejes principales, un dato que –en algunos casos, como dijimos– nos trae a la memoria más el culto aymara (Bouysse-Cassagne 1987) que las *Instrucciones* del cardenal Borromeo. En la Figura 4 se muestran las orientaciones del grupo completo de iglesias y sus emplazamientos en la región estudiada, lo que nos permite analizar, caso por caso, la posible preponderancia del paisaje terrestre por encima del celeste.

**Conclusiones**

Pese a tener un origen común, como respuesta evangelizadora para multitudes de trajinantes que pasaban por los nuevos poblados diseminados a lo largo de la ruta de la plata, las iglesias andinas patrimoniales de la extensa y difícil de transitar región de Arica y Parinacota muestran una cierta diversidad. Esto es visible en varios planos, ya sea en el de sus materialidades y sistema constructivo, como en lo que respecta a sus ornamentos, pinturas murales y arquitectura general, elementos que muestran la intención y dedicación de las comunidades originales que las construyeron. Por otra parte, los emplazamientos geográficos elegidos, en los valles bajos, precordillera y altiplano, y los paisajes diversos en donde estas construcciones se hayan inmersas, proveen datos adicionales que nos ayudan a pensar en las características globales de los templos cristianos.

Con la idea de sumar nuevos elementos para reflexionar sobre estas iglesias, en este trabajo hemos estudiado sus orientaciones espaciales precisas, apoyándonos en los métodos usuales de la arqueoastronomía. Nuestros resultados muestran que, a diferencia de lo que se encuentra en los estudios llevados a cabo con iglesias antiguas europeas, por mencionar un ejemplo, en las iglesias andinas patrimoniales medidas aquí no se siguió un único patrón de orientación determinante en toda la región. Sin embargo, hemos hallado que casi la mitad de las iglesias estudiadas se orienta dentro del rango solar, con una proporción dominante en aquellas que presentan su altar hacia el poniente. Hemos también señalado unos pocos casos notables en los que las orientaciones de los templos parecen obedecer más a la ubicación de elementos distintivos del paisaje terrestre – volcanes como cerros tutelares, por ejemplo– que a la salida o puesta del Sol en fechas relevantes para la advocación particular de las iglesias.

Un caso particular que vale la pena discutir brevemente aquí es el de la iglesia San José de Parcohaylla. Si dejamos de lado la incertidumbre en la medición de la altura del horizonte (que puede rondar aproximadamente unos pocos grados), esta iglesia se orienta en forma aproximadamente equinoccial, alineando su eje hacia el poniente (y con declinación astronómica muy cercana a cero; véase la Tabla 1). La fecha de su fiesta patronal es el 19 de marzo, a pocos días del equinoccio, por lo que podemos deducir que probablemente esta iglesia fue ex profeso orientada hacia la puesta del Sol en el día de San José. Sin embargo, este es un ejemplo aislado; el resto de las iglesias patrimoniales no parece obedecer esta regla.

Otro caso digno de señalar es el de la iglesia Santiago Apóstol de Airo, ubicada en el extremo norte de la región y en las inmediaciones del volcán Tacora. A diferencia de lo ya mencionado antes para la iglesia Virgen del Carmen de Tacora, en la de Airo se da un caso que podríamos llamar *mixto*. La alineación de su eje principal entra en el rango solar (pues su declinación es -12.5°) y, al mismo tiempo, su acimut (257.5°) apunta al poniente a solo 7.5° de la dirección del volcán Tacora que domina el paisaje (cuyo acimut, medido desde la iglesia, es de 265°). La diferencia angular entre los acimutes –menor que el ángulo que subtiende el puño de una mano, cuando se mantiene el brazo extendido– es lo suficientemente pequeña como para pensar que probablemente ese imponente monte nevado de referencia estuvo *también* en la mira –y entre las intenciones de alineación– de los constructores de esta iglesia.

A diferencia de lo hallado en otros sitios de América, por ejemplo en la región de Xochimilco, hacia el sur de la Ciudad de México (ver, por ejemplo, Zimbrón Romero 1992, quien estudia las "cruces punteadas" de Santa Cruz Acalpixcan), en las regiones investigadas en nuestro trabajo no hemos hallado referencias a marcadores de horizonte, es decir, piedras talladas u otros artefactos que apuntasen a volcanes o cerros conspicuos del paisaje. En sus estudios, el mencionado autor pudo verificar que una cruz punteada (un petroglifo en forma de cruz sobre piedra basáltica) sirvió para alinear antiguas construcciones prehispánicas de acuerdo a la ubicación del Sol en el horizonte en momentos específicos del calendario agrícola. Posteriormente, sobre estas construcciones se edificaron las actuales iglesias coloniales del pueblo de Santa Cruz Acalpixcan, cuyas advocaciones coinciden muy aproximadamente con las festividades cristianas que toman lugar en esos días (la "Santa Cruz" el 3 de mayo y "San Salvador" el 6 de agosto; ver Zimbrón Romero 1992:72). Sin embargo, como mencionamos, en nuestro trabajo esto no se replica.

Somos conscientes de que el grupo de iglesias patrimoniales de esta región –aun cuando su número fue el adecuado para un estudio estadístico como el realizado aquí– no agota el trabajo que se debería abordar en un futuro próximo. La región de Arica y Parinacota está culturalmente muy emparentada con zonas del oeste boliviano y del sur del Perú, no solo en lo que a geografía se refiere, sino –y especialmente– en lo relacionado con las costumbres y religiosidad de sus pobladores, tanto actuales como de los siglos pasados (Díaz et al. 2010; Díaz et al. 2012).

Recientemente hemos ejecutado un primer estudio exploratorio de las orientaciones de algunas iglesias antiguas que rodean a la ciudad de Arequipa, Perú, junto a otras ubicadas en la margen izquierda del Valle del Colca (Gutiérrez et al. 1986). Un análisis preliminar de las mediciones de estas nuevas iglesias del sur peruano indica que, muy probablemente, los patrones de orientación hallados en el norte de Chile quizá se repliquen en los alrededores de Arequipa. Hemos verificado, por ejemplo, que varias iglesias del Colca se orientan en el rango solar y, en particular, una de ellas, Nuestra Señora de la Asunción de Chivay, además de orientarse a levante, tiene la propiedad de que, al salir del templo por su puerta principal, el visitante se enfrenta con el volcán Hualca Hualca con una precisión de solo cinco grados de acimut respecto del eje de la iglesia.[12]

En resumen, pensamos que el estudio de las orientaciones espaciales precisas de las iglesias andinas antiguas que aquí hemos iniciado puede extenderse a regiones vecinas y, de esa manera, completarse la caracterización de estas construcciones emblemáticas del territorio surandino, como un complemento necesario a la investigación y análisis histórico y cultural de los templos.

**Listado de tablas**

Tabla 1. Orientaciones de las iglesias patrimoniales de la región de Arica y Parinacota, obtenidas en nuestra misión de marzo de 2015. Para cada construcción, la tabla muestra la ubicación, la identificación (nombre, fecha de primera mención, cuando está disponible, y fecha más probable de construcción de la iglesia actual), la latitud y longitud geográficas (L y l), el acimut astronómico (a) tomado a lo largo del eje del edificio en dirección al altar (redondeado al ½° de error), la altura angular del horizonte (h) en esa dirección (0 B significa horizonte bloqueado, se toma h=0°), y la declinación resultante correspondiente (δ).

Tabla 1. Orientaciones de las iglesias de la región de Arica y Parinacota.

| Ubicación | Nombre / fecha | L (°/') Sur | l (°/') Oeste | a (°) | h (°) | δ (°) | Fecha santo / orientación |
|---|---|---|---|---|---|---|---|
| (1) Azapa | San Miguel (1618; fines s. XIX) | 18/31 | 70/11 | 36½ | 15 | 42 | 29 Sep / ---- |
| (2) Poconchile | San Jerónimo (1618; fines s. XIX) | 18/27 | 70/04 | 156 | 15 | -66½ | 30 Sep / ----- |
| (3) Chitita | Virgen del Carmen (fines s. XIX) | 18/50 | 69/41 | 299½ | 11 | 24½ | 16 Jul / ----- |
| (4) Guañacagua | San Pedro (fines s. XIX) | 18/49 | 69/43 | 78½ | 14 | 6½ | 29 Jun / 7 Abr – 6 Sep |
| (5) Sucuna | San Antonio de Padua (fines s. XIX) | 18/51 | 69/27 | 275½ | 0 | 5½ | 13 Jun / 4 Abr – 10 Sep |
| (6) Saguara | La Santa Cruz (último tercio s. XIX) | 18/54 | 69/30 | 266½ | 1 | -3½ | 3 Mayo / 12 Mar – 2 Oct |
| (7) Pachica | San José (1618; comienzos s. XVIII) | 18/55 | 69/37 | 76 | 9 | 10½ | 19 Mar / 18 Abr – 26 Ago |
| (8) Esquiña | San Pedro (1618; comienzos s. XVIII) | 18/56 | 69/32 | 108½ | 15 | -21½ | 29 Jun / 14 Ene – 30 Nov |
| (9) Aico | San Antonio de Padua (fines s. XIX) | 18/48 | 69/28 | 69½ | 6 | 18 | 13 Jun / 12 Mayo – 2 Ago |
| (10) Parcohaylla | San José (fines s. XIX) | 18/53 | 69/13 | 271 | 0 B | 1½ | 19 Mar / 24 Mar – 20 Sep |
| (11) Mulluri | Virgen de la Natividad (fines s. XIX) | 19/01 | 69/10 | 244 | 5 | -26 | 8 Sep / ----- |
| (12) Codpa | San Martín de Tours (1618; comienzos s. XVIII) | 18/50 | 69/45 | 70½ | 13 | 14½ | 11 Nov / 29Abr – 15 Ago |
| (13) Timar | San Juan Bautista (1618; comienzos s. XVIII) | 18/45 | 69/41 | 53½ | 8 | 31½ | 24 Jun / ----- |
| (14) Cobija | San Isidro Labrador (segunda mitad s. XIX) | 18/44 | 69/35 | 255½ | 5 | -15 | 15 Mayo / 8 Feb – 4 Nov |
| (15) Timalchaca | Virgen de los Remedios (fines s. XIX) | 18/41 | 69/25 | 309½ | -1 | 38 | 21 Nov / ----- |
| (16) Tignamar | Virgen Asunción (1618; fines s. XIX) | 18/35 | 69/29 | 101½ | 6 | -12½ | 15 Ago / 17 Feb – 26 Oct |
| (17) Chapiquiña | San José Obrero (mediados del s. XX) | 18/24 | 69/32 | 63½ | 12 | 21 | 1 Mayo / 27 May – 18 Jul |
| (18) Pachama | San Andrés Apóstol (s. XVIII) | 18/26 | 69/31 | 19½ | 10 | 57 | 30 Nov / ----- |
| (19) Saxamar | Santa Rosa de Lima (1618; s. XX) | 18/33 | 69/30 | 320½ | 3½ | 46 | 30 Ago / ----- |
| (20) Belén | Santiago Apóstol (1618; comienzos s. XVIII) | 18/28 | 69/31 | 188½ | 16 | -81½ | 25 Jul / ----- |
| (21) Belén | Virgen de la Candelaria (segunda mitad s. XVIII) | 18/28 | 69/31 | 162½ | 19 | -73 | 2 Feb / ----- |
| (22) Socoroma | San Francisco de Asís (1618; fines s. XIX) | 18/16 | 69/36 | 15½ | 4½ | 63½ | 4 Oct / ----- |
| (23) Putre | Virgen Asunción (1618; fines s. XIX) | 18/12 | 69/34 | 352½ | 10 | 62 | 15 Ago / ----- |
| (24) Tacora | Virgen del Carmen (s. XVIII) | 17/46 | 69/43 | 289½ | 4 | 17½ | 16 Jul / 10 May – 4 Ago |
| (25) Airo | Santiago Apóstol (comienzos s. XX) | 17/43 | 69/39 | 257½ | 3 | -12½ | 25 Jul / 16 Feb – 27 Oct |
| (26) Chapoco | San Martín de Tours (mediados s. XX) | 17/43 | 69/34 | 203 | 6 | -65 | 11 Nov / ----- |
| (27) Putani | Virgen Inmaculada Concepción (s. XX) | 17/44 | 69/32 | 207 | 3½ | -60 | 8 Dic / ----- |
| (28) Pucoyo | Virgen del Rosario (comienzos s. XX) | 17/46 | 69/25 | 253 | 6½ | -18 | 3 Oct / 30 Ene – 13 Nov |
| (29) Cosapilla | Virgen del Rosario (s. XVIII) | 17/46 | 69/25 | 219 | 13 | -52 | 3 Oct / ----- |
| (30) Guacollo | Santa Rosa de Lima (comienzos s. XX) | 17/46 | 69/21 | 234 | 4 | -35½ | 30 Ago / ----- |
| (31) Parinacota | Virgen de la Natividad (s. XVIII) | 18/12 | 69/16 | 37 | 2½ | 48½ | 8 Sep / ----- |
| (32) Ungallire | Virgen del Carmen (fines s. XIX) | 18/12 | 69/16 | 249 | 3 | -20½ | 16 Jul / 18 Ene – 25 Nov |
| (33) Caquena | Santa Rosa de Lima (fines s. XIX) | 18/03 | 69/12 | 268 | 8 | -4 | 30 Ago / 10 Mar – 4 Oct |
| (34) Chucuyo | La Santa Cruz (fines s. XIX) | 18/13 | 69/18 | 270 | 0 B | 0 | ?? Mayo / 22 Mar – 23 Sep |
| (35) Choquelimpie | Virgen Asunción (1618; fines s. XIX) | 18/19 | 69/16 | 197 | 20 | -74 | 15 Ago / ----- |
| (36) Ancuta | Virgen Inmaculada Concepción (fines s. XIX) | 18/27 | 69/12 | 252 | 1 | -17 | 8 Dic / 1 Feb – 11 Nov |
| (37) Guallatire | Virgen Inmaculada Concepción (1873) | 18/30 | 69/09 | 218 | 1 | -49 | 8 Dic / ----- |
| (38) Churiguaya | Virgen de la Candelaria (fines s. XIX) | 18/21 | 69/11 | 232 | 2 | -36½ | 2 Feb / ----- |

# Notas

[1] Una muy útil secuencia cronológica de los diferentes períodos correspondientes a la región que estamos considerando en este trabajo, puede verse en (Chacama y Díaz 2011:44).

[2] Otros casos de nuevos poblados ubicados en cercanías de los antiguos pueblos prehispánicos de la precordillera de Arica son señalados en (Chacama 2009).

[3] En términos generales, la arquitectura religiosa que empezó a materializarse en el territorio surandino, incluido los Altos de Arica y sus singularidades de entorno, se afianzó en las siguientes características. En primer lugar, los atisbos arquitectónicos provenientes del siglo XVII no fueron pretenciosos en lo funcional y resultaron modestos y sencillos en los aspectos constructivos. Se escogieron plantas uniespaciales alargadas, ábsides ochavados, muros de piedra sin cantear, cubiertas a dos aguas con soportes de par y nudillo, arcos torales de separación y techumbres de paja. En las áreas exteriores, aparecieron las cruces catequísticas, los atrios arqueados, las torres-campanario exentas, las capillas misereres y posas. En los entornos circundantes a las fábricas religiosas se hicieron notar con claridad las demandas indígenas materializadas en el uso sacralizado de los espacios abiertos, formalizándose la cohabitación entre las ideologías morfológicas europeas e indígenas. Los templos estudiados en este trabajo son herederos de esta tradición (Fundación Altiplano 2012).

[4] En base a un riguroso estudio histórico-cultural y arquitectónico, se ha seleccionado un grupo de aproximadamente cuarenta iglesias patrimoniales que son las que hemos incluido en este trabajo. En la región hay por lo menos otras cuarenta construcciones religiosas adicionales, que no hemos tenido en cuenta por tratarse de capillas altiplánicas de época muy reciente, aproximadamente de bien avanzado el siglo XX.

[5] La declinación es una coordenada ecuatorial celeste y se calcula a partir del acimut, la altura angular del horizonte y la latitud del sitio de cada iglesia. Tiene en cuenta las correcciones en la altura angular del horizonte debidas a la refracción atmosférica. La declinación del Sol, por ejemplo, representa la distancia angular a la que se halla el Sol del ecuador celeste, y varía entre aproximadamente -23.5° (valor correspondiente al día del solsticio de Diciembre) y +23.5° (correspondiente al día del solsticio de Junio).

[6] En particular, divisamos calvarios y pequeñas apachetas o, más apropiadamente, *markas* (Muñoz y Briones 1998:56), algunas de ellas con cruces, en muchos montes visibles desde las iglesias y tomamos nota de la ubicación de montañas o volcanes prominentes en la cercanía, así como también de la presencia de ríos y otros aspectos topográficos relevantes. Discutiremos algunas de estas observaciones en las líneas que siguen.

[7] La excepción fue la iglesia del poblado de Choquelimpie, a la que no pudimos acceder por hallarse dentro de los confines infranqueables de un viejo mineral, cuya orientación debimos estimar a partir del análisis de fotos satelitales.

El paisaje circundante al templo, incluyendo la medida de la altura de su horizonte en la dirección del eje, fue verificado a partir de fotos obtenidas previamente por colegas de la Fundación Altiplano.

[8] Son seis si solo consideramos el acimut. En realidad la medida relevante es la declinación astronómica, pues con el acimut por sí solo no se tiene en cuenta la altura angular del horizonte en la dirección del eje de la iglesia. Si consideramos la declinación (calculada a partir del acimut y de la mencionada altura angular), son siete las iglesias que entran en el rango solar. La iglesia que se agrega a la lista es la de Chapiquiña (ver Tabla 1).

[9] Ver también la comparación con las iglesias históricas de la isla de Lanzarote en (Gangui et al. 2016).

[10] Muchas tumbas sobreelevadas (torrecillas funerarias o *chullpas*) de regiones vecinas se hallan orientadas hacia cerros prominentes y volcanes (Pärssinen 2005).

[11] Al inspeccionar esta capilla, el Arq. Patricio Parra nos hizo notar que, aparentemente, hubo una gran abertura -hoy tapada- en el muro testero, donde hoy se halla el altar. En caso de haber sido ésta la puerta de acceso original, la orientación del eje de este pequeño templo habría estado dirigida (casi directamente) a los payachatas. Sin embargo, una explicación más prosaica es que frecuentemente se hace una abertura en el muro testero para que la luz del día ilumine a la Custodia, como sucede en otras iglesias de la región, la de Pachama por ejemplo; a veces esas aberturas luego se agrandan como ventanas, como sucede en las iglesias de Socoroma y de Esquiña.

[12] Más elementos para reflexionar sobre el Valle del Colca como paisaje cultural pueden verse en Mujica Barreda y de la Vera Cruz 2001.

# LA ORIENTACIÓN DE LAS IGLESIAS ANDINAS DE LA REGIÓN DE ARICA Y PARINACOTA, CHILE: UNA APROXIMACIÓN ARQUEOASTRONÓMICA

*Alejandro Gangui*, *Ángel Guillén* y *Magdalena Pereira*

**Figuras**

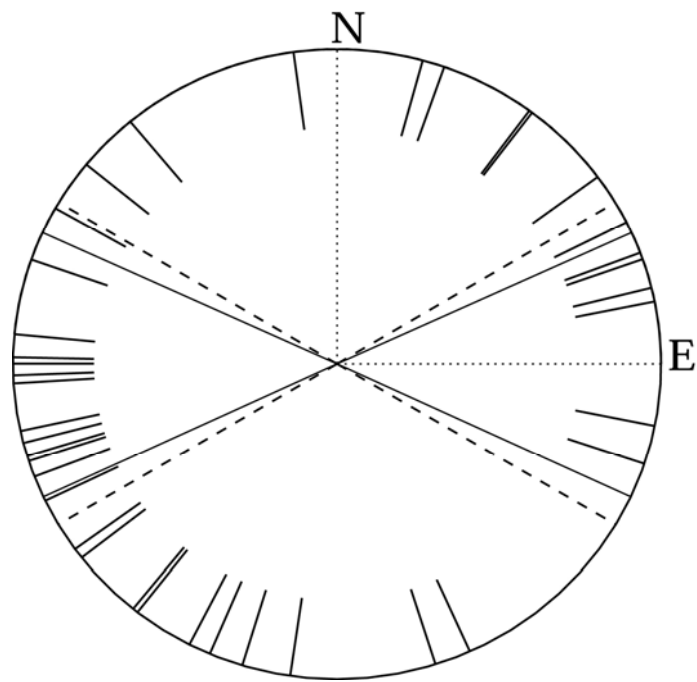

Figura 1. Diagrama de orientación para las iglesias y capillas andinas patrimoniales de la región de Arica y Parinacota, obtenido a partir de los datos de la Tabla 1. Se puede ver que existe una gran diversidad de orientaciones; pese a esto, un número significativo (un poco menos de la mitad) sigue un patrón de orientación dentro del rango solar.

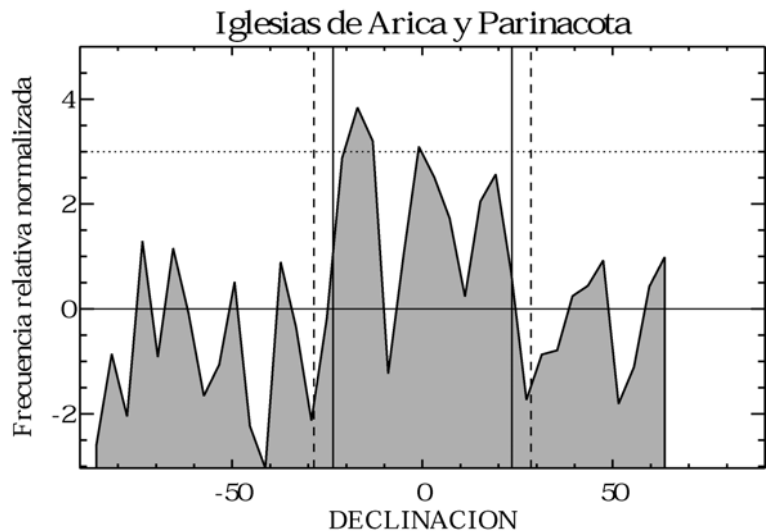

Figura 2. Histograma de declinaciones para las iglesias y capillas andinas patrimoniales obtenido de los datos de marzo de 2015. Se encuentran dos picos estadísticamente significativos, por encima del nivel 3σ, y no muy notorios (en total, son tres picos dentro del rango solar). Las líneas verticales continuas representan las declinaciones correspondientes a las posiciones extremas del Sol en los solsticios (-23.5° y +23.5°), mientras que las verticales discontinuas representan lo propio para la Luna en los lunasticios mayores. La frecuencia de declinaciones presente fuera del rango lunisolar no presenta picos significativos y nos sugiere que un gran número de construcciones fueron orientadas siguiendo lineamientos ajenos a los canónicos.

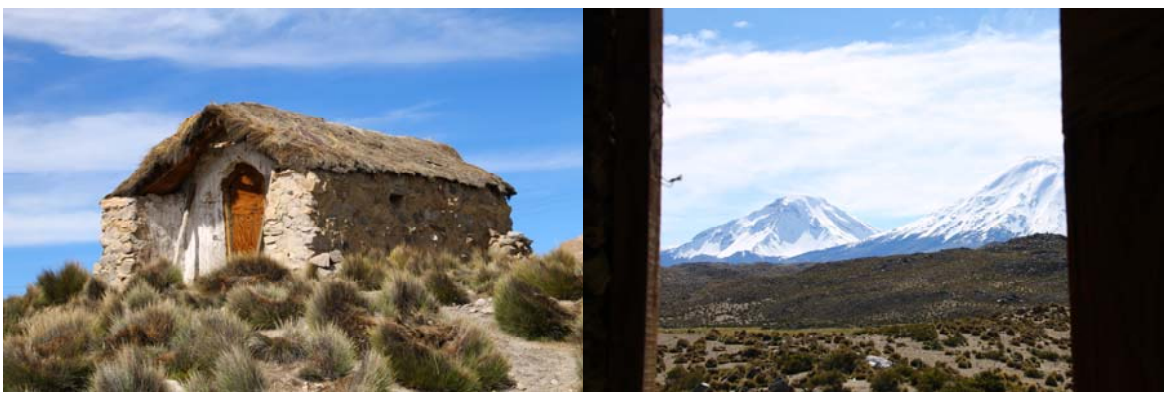

Figura 3. La capilla de Ungallire en su emplazamiento en cercanías del poblado de Parinacota (imagen izquierda) y la visión de los nevados de payachatas al salir del templo (imagen derecha), en dirección aproximadamente igual al de su eje, pero en sentido opuesto al del altar.

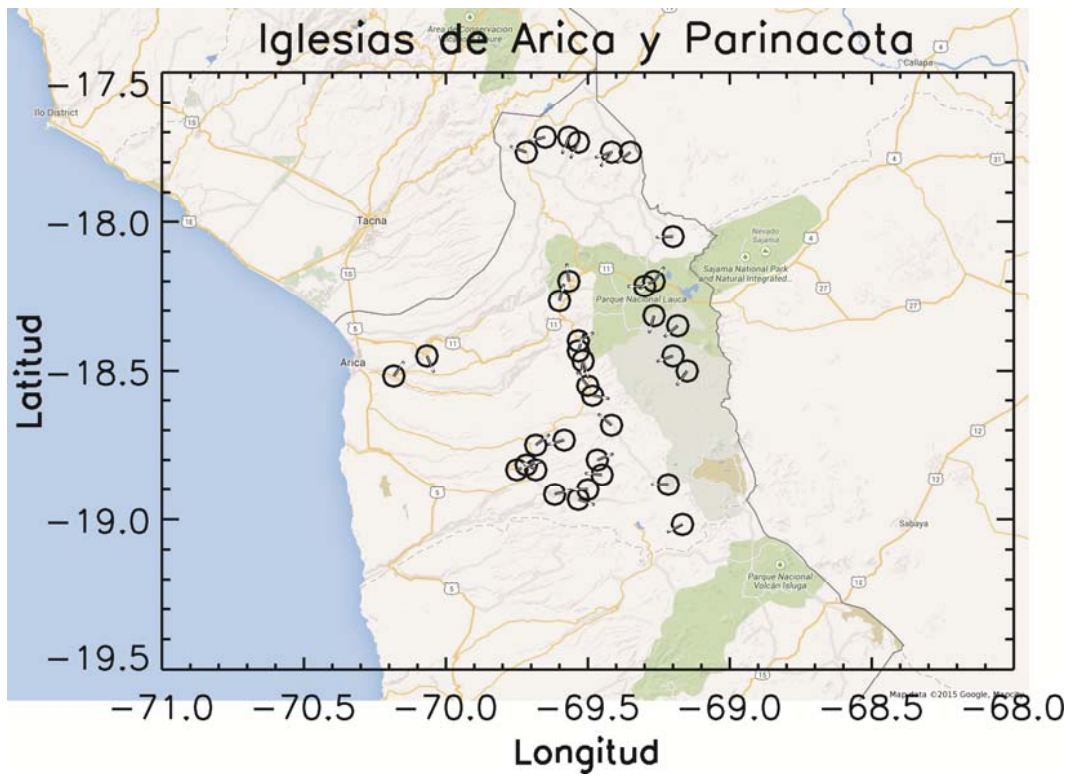

Figura 4. Mapa con la ubicación geográfica de la totalidad de las iglesias y capillas andinas patrimoniales medidas (señaladas con círculos), junto con la orientación del eje de las construcciones en dirección al altar (flechas, orientadas de acuerdo a los acimutes consignados en la Tabla 1). En la parte superior del mapa, en las inmediaciones del poblado de Cosapilla, dos iglesias geográficamente muy próximas (Pucoyo y Cosapilla) poseen orientaciones diferentes, lo que explica la presencia allí de un único círculo con dos flechas. Lo mismo sucede para un par de iglesias (Parinacota y Ungallire) en el interior del Parque Nacional Lauca, y para las iglesias Santiago Apóstol y Virgen de la Candelaria del poblado de Belén, hacia el centro del mapa. Imagen de los autores sobre un mapa cortesía de Google Maps.